\documentclass[sigconf]{acmart}
\usepackage{enumitem}
\usepackage{multirow}
\usepackage[capitalise]{cleveref}

\makeatletter
\def\@ACM@checkaffil{
    \if@ACM@instpresent\else
    \ClassWarningNoLine{\@classname}{No institution present for an affiliation}%
    \fi
    \if@ACM@citypresent\else
    \ClassWarningNoLine{\@classname}{No city present for an affiliation}%
    \fi
    \if@ACM@countrypresent\else
        \ClassWarningNoLine{\@classname}{No country present for an affiliation}%
    \fi
}
\makeatother
\AtBeginDocument{%
  \providecommand\BibTeX{{%
    \normalfont B\kern-0.5em{\scshape i\kern-0.25em b}\kern-0.8em\TeX}}}

\setcopyright{acmcopyright}
\copyrightyear{2018}
\acmYear{2018}
\acmDOI{XXXXXXX.XXXXXXX}

\acmConference[Conference acronym 'XX]{Make sure to enter the correct
  conference title from your rights confirmation emai}{June 03--05,
  2018}{Woodstock, NY}
\acmPrice{15.00}
\acmISBN{978-1-4503-XXXX-X/18/06}

\begin{document}

\title{TBIN: Modeling Long Textual Behavior Data for CTR Prediction}

\author{Shuwei Chen}
\authornote{Co-first authors with equal contributions.}
\email{chenshuwei04@meituan.com}
\affiliation{Meituan}

\author{Xiang Li}
\authornotemark[1]
\email{lixiang172@meituan.com}
\affiliation{Meituan}

\author{Jian Dong}
\email{dongjian03@meituan.com}
\affiliation{Meituan}

\author{Jin Zhang}
\email{zhangjin11@meituan.com}
\affiliation{Meituan}

\author{Yongkang Wang}
\email{wangyongkang03@meituan.com}
\affiliation{Meituan}

\author{Xingxing Wang}
\email{wangxingxing04@meituan.com}
\affiliation{Meituan}
\renewcommand{\shortauthors}{S. Chen and X. Li, et al.}

\begin{abstract}

Click-through rate (CTR) prediction plays a pivotal role in the success of recommendations.
Inspired by the recent thriving of language models (LMs), a surge of works improve prediction by organizing user behavior data in a \textbf{textual} format and using LMs to understand user interest at a semantic level.
While promising, these works have to truncate the textual data to reduce the quadratic computational overhead of self-attention in LMs.
However, it has been studied that long user behavior data can significantly benefit CTR prediction.
In addition, these works typically condense user diverse interests into a single feature vector, which hinders the expressive capability of the model.
In this paper, we propose a \textbf{T}extual \textbf{B}ehavior-based \textbf{I}nterest Chunking \textbf{N}etwork (TBIN), which tackles the above limitations by combining an efficient locality-sensitive hashing algorithm and a shifted chunk-based self-attention.
The resulting user diverse interests are dynamically activated, producing user interest representation towards the target item.
Finally, the results of both offline and online experiments on real-world food recommendation platform demonstrate the effectiveness of TBIN.

\end{abstract}



\maketitle

\section{Introduction}

The accuracy of click-through rate (CTR) prediction~\cite{wide_deep, deepfm, dcn, dcin, li2023context}, which estimates the probability of a user clicking on a target item, has a direct impact on user satisfaction and platform revenue for recommendations.
Inspired by the recent success of language models (LMs)~\cite{gpt3, instructgpt, glm, llama, llama2}, numerous works have introduced LMs into recommendation tasks and achieved impressive results, including
sequential recommendation~\cite{idasr, unis_rec, recformer}, CTR prediction~\cite{ctr_bert, ctrl}, multitask learning~\cite{m6_rec, p5}, etc.
In these works, the model input such as user behavior data (composed of user interacted items) is converted into \textbf{textual} description and fed into LMs~\cite{bert,roberta} to extract user interest at a semantic level.

Although semantic information improves prediction to some extent, there are some limitations to these works.
On the one hand, they require truncating the textual behavior data to reduce the quadratic computational overhead of self-attention~\cite{attention}, which is widely adopted as the building block of LMs.
However, previous studies~\cite{dien, mimn, sim, hpmn} have demonstrated that incorporating long user behavior data into the models helps extract rich user interest and benefits CTR prediction.
On the other hand, user interests are diverse~\cite{din} and these works condense user interests into a single feature vector, such as the [\emph{CLS}] token in transformer, hindering the expressive capability of the model.
This limitation is particularly evident in scenarios where users have multiple interests or when their interests are constantly evolving.
In such cases, a single feature vector fails to capture the complexity and nuance of the user's interests, resulting in a less accurate and personalized recommendation.

To mitigate the aforementioned limitations, this paper proposes a \textbf{T}extual \textbf{B}ehavior-based \textbf{I}nterest chunking \textbf{N}etwork (TBIN).
TBIN first converts long textual behavior data into an embedding sequence by feeding the textual description of each item into a pre-trained BERT~\cite{bert}, and uses transformer blocks to model the correlation between different item embeddings.
To avoid the $\mathcal{O}(L^2)$ complexity of self-attention, where $L$ is the length of the embedding sequence, TBIN uses an efficient Locality-Sensitive Hashing (LSH) algorithm~\cite{projection, reformer} to evenly partition the sequence into chunks containing similar embeddings.
TBIN then extracts user diverse interests by applying self-attention separately in each chunk, changing the complexity to $\mathcal{O}(L~\text{log}~L)$.
Furthermore, considering that adjacent chunks are highly correlated, we propose a Shifted Chunk-based Self-Attention (SC-SA) inspired by Swin Transformer~\cite{swin} to allow for cross-chunk connection.
Finally, TBIN conducts target attention to dynamically activate user diverse interests, producing user interest representation towards the target item, which will be concatenated with other features to predict CTR.

We conduct offline and online experiments on real-world food recommendation platform.
The experimental results demonstrate the effectiveness of the proposed TBIN.
\section{Preliminaries}
\label{sec:pre}

\begin{figure}[t]
  \centering
  \includegraphics[width=\linewidth]{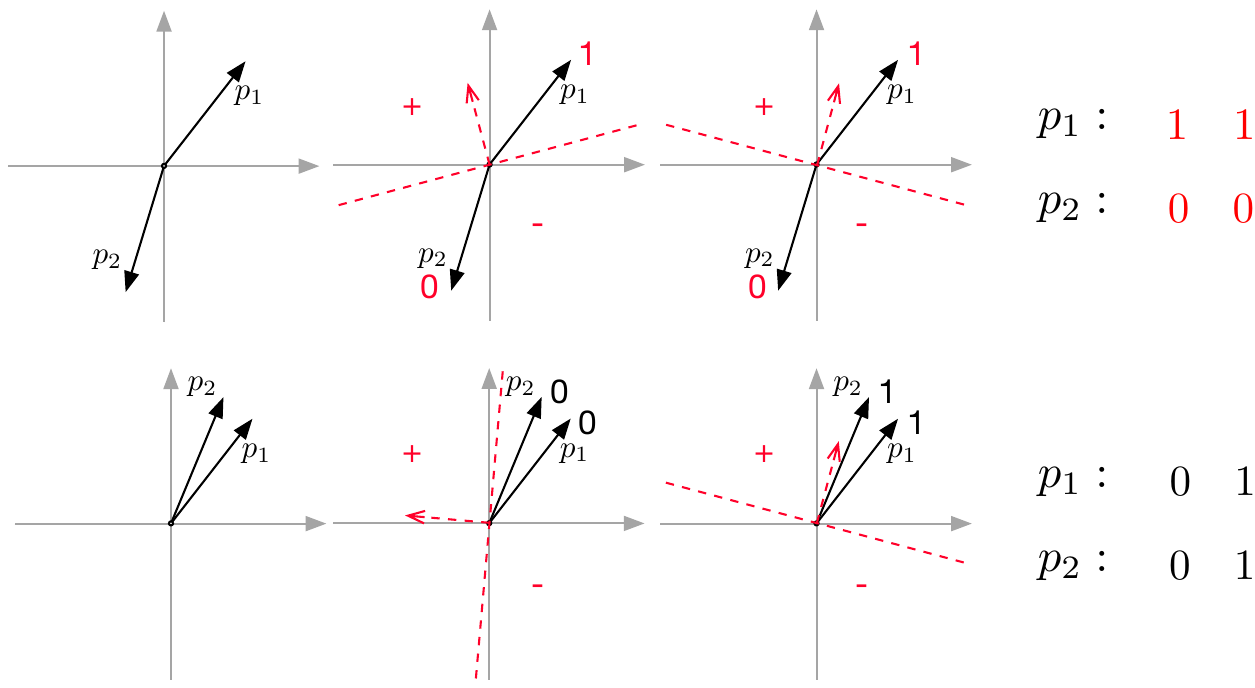}
  \caption{
  Random projection is a LSH algorithm.
  The \text{\color{red}red} random vectors split 2D-planes into positive (\text{\color{red}+}) and negative (\text{\color{red}-}) half-planes.
  Vectors projected onto the random vector with a positive value are encoded as 1, while those with a negative value are encoded as 0.
  By executing random projection multiple times, the hash codes of dissimilar vectors are likely to be different, while the hash codes of similar vectors are likely to be the same.}
  \label{fig:lsh}
\end{figure}

\begin{itemize}[leftmargin=*]
    \item \textbf{Problem Setup.}
    TBIN takes target item features $x^t$, user profile features $x^u$, and long textual behavior data $x^b$ as inputs to predict the user's CTR on the target item.
    Specifically, the behavior data $x^b=\{b_1, b_2,\dots, b_L\}$ contains $L$ sentences, where $b_i = \{w^i_1, w^i_2, \dots, w^i_n\}$ contains $n$ words describing the $i$-th item that the user interacted with.
    Note that $L$ is large.

    \item \textbf{Self-Attention (SA).} 
    SA~\cite{attention} is capable of capturing overall dependency among input tokens.
    Yet, when applied to $L$ tokens, it multiplies the query and key matrix to get a weight matrix with size $L \times L$, followed by a softmax function.
    This leads to a quadratic complexity and makes SA infeasible when $L$ is large.

    \item \textbf{Locality-Sensitive Hashing (LSH).}
    The LSH algorithm has a high probability of allocating the same hash codes to similar vectors, making it suitable for fast and approximate searching of similar vectors.
    This paper follows previous works~\cite{reformer, eta} to use random projection~\cite{projection} as the LSH algorithm, to improve SA efficiency when modeling long textual behavior data (described in \cref{sec:method}),
    and \cref{fig:lsh} provides visual explanation in a 2D plane that random projection is a LSH algorithm. 
\end{itemize}

\section{Methodology}
\label{sec:method}

\begin{figure}[t]
  \centering
  \includegraphics[width=\linewidth]{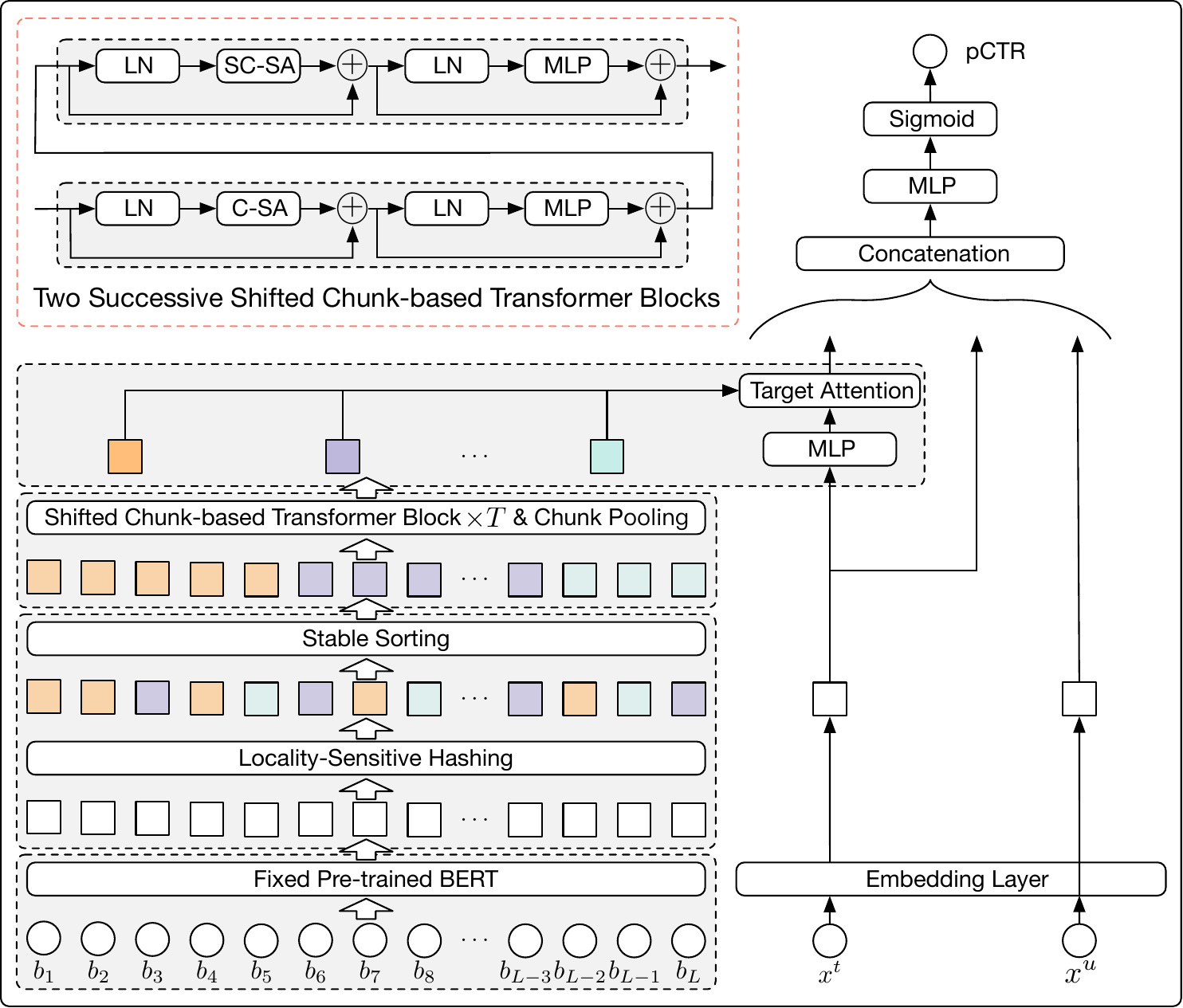}
  \caption{The architecture of TBIN and the structure of two successive shifted chunk-based transformer blocks.}
  \label{fig:arch}
\end{figure}

\cref{fig:arch} illustrates the overall architecture of the proposed Textual Behavior-based Interest Chunking Network (TBIN).
To begin with, we input the textual descriptions of each item in the long behavior data into a pre-trained BERT to obtain an embedding sequence. 
As the parameters in BERT are fixed, computations in this step can be pre-executed and stored to improve model efficiency.
Then, we use a locality-sensitive hashing algorithm to  allocate hash codes for the embeddings.
Embeddings with the same hash code (colored the same) are considered similar, and we sort them together.
Next, several shifted chunk-based transformer blocks are applied to extract user diverse interests, which will be dynamically activated by target attention to obtain the final user interest representation w.r.t. the target item.
This representation is concatenated with other feature embeddings to predict CTR.
For clarity, we continue to use $x^b, x^t, x^u$ to denote the embeddings of behavior data/target item/user, and provide more details in the following sections.

\subsection{Locality-Sensitive Hashing-based Sorting.}
\label{sec:lsh}
Despite the powerful sequence modeling capability of transformer,  the $\mathcal{O}(L^2)$ complexity of self-attention (SA) makes it inefficient for modeling \textbf{long} sequence of size $L$.
This paper follows previous works~\cite{reformer} to use locality-sensitive hashing (LSH) algorithm to improve transformer efficiency.
The core idea is that in SA, the softmax function is dominated by a few elements with large weights~\cite{attention}.
Therefore, each query actually only needs to focus on the keys that contribute the most to it (i.e., the keys that are most similar to the query) in order to avoid unnecessary computations, and the LSH algorithm can efficiently search for similar vectors.

Specifically, we choose random projection~\cite{projection} as the implementation of LSH.
Given behavior data embeddings $x^b \in \mathbb{R}^{L \times d_b}$, we generate a random matrix $\mathbf{R} \in \mathbb{R}^{d_b \times m}$ and use the following formula to obtain a hash code matrix :

\begin{equation}
    \mathbf{H} = \mathbb{I}(x^b \mathbf{R})
\end{equation}
where the value of the indicator function $\mathbb{I}(x)$ is 1 when x is positive, and 0 otherwise. $\mathbf{H} \in \{0,1\}^{L \times m}$ is the hash code matrix, and each row corresponds to an $m$-bit hash code for an embedding.

We use stable sorting based on hash codes to partition embeddings into buckets, where embeddings within the same bucket have the same hash code and are similar to each other.
We will describe how to efficiently extract user diverse interests in the next section.

\begin{figure}[t]
  \centering
  \includegraphics[width=\linewidth]{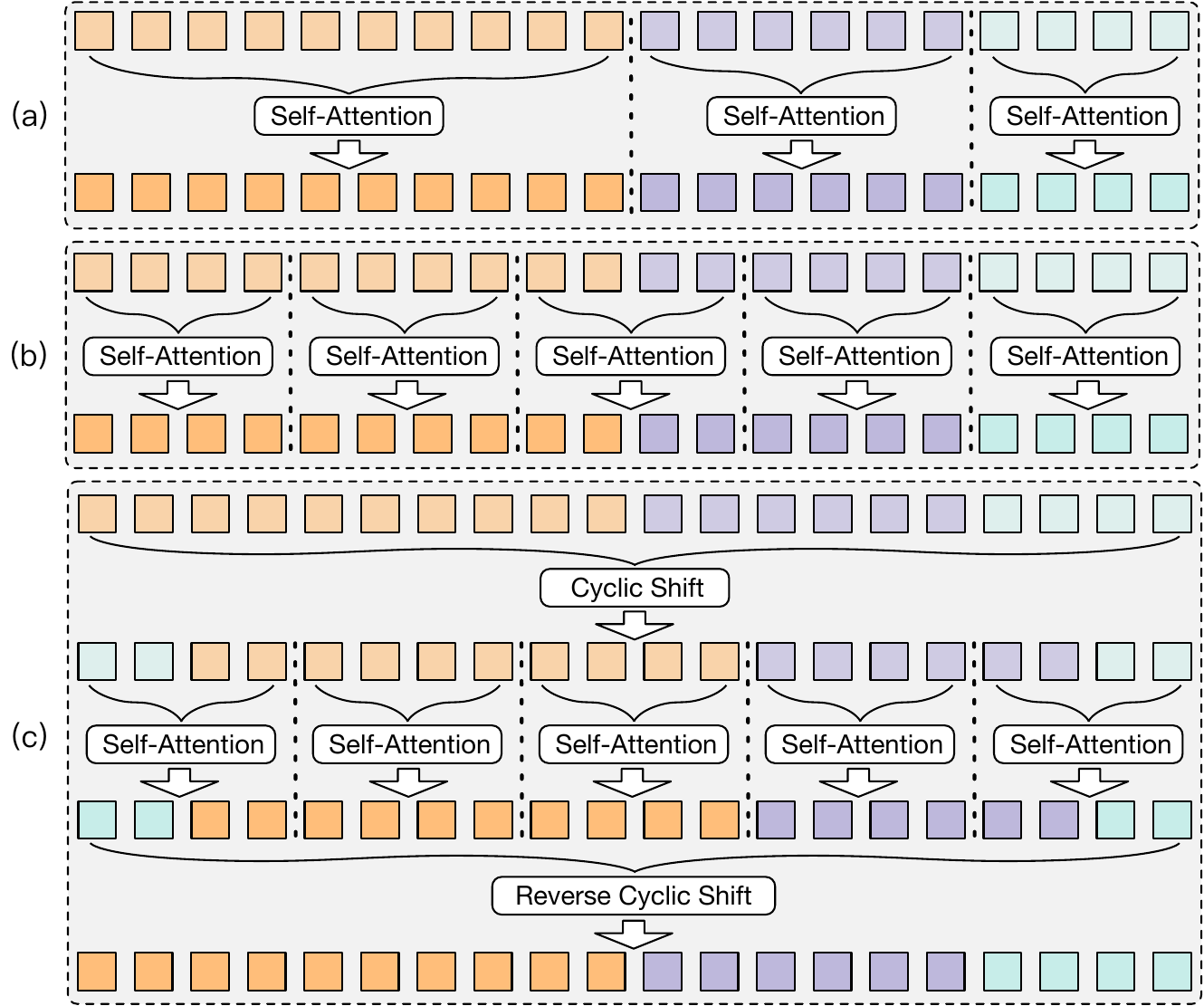}
  \caption{Illustration of different self-attention schemas.
  (a) Bucket-based Self-Attention (B-SA); (b) Chunk-based Self-Attention (C-SA); (c) Shifted Chunk-based Self-Attention (SC-SA).
  Embeddings in the same bucket are colored the same, and buckets or chunks are separated by vertical dashed lines.
  }
  \label{fig:sa}
\end{figure}

\subsection{Shifted Chunk-based Self-Attention}

Similar behavior embeddings within a bucket indicate a certain type of user interest, and performing self-attention separately in each bucket can learn user diverse interests (as shown in \cref{fig:sa}(a)).
However, this self-attention schema poses two issues: 1) the different sizes of buckets make it difficult to parallelize the computations, and 2) if a bucket contains a large number of embeddings, the model complexity may not be significantly reduced.
We further improve model efficiency with chunk-based self-attention (C-SA).
As shown in \cref{fig:sa}(b), the buckets are evenly divided into fine-grained chunks, and self-attention is calculated for each chunk separately.
In this way, the complexity of self-attention is reduced to $\mathcal{O}(L~\text{log}~L)$.
Yet, this schema introduces another limitation:
the highly correlated embeddings in the same bucket are isolated and independently modeled, and their internal relationships are not utilized, which may result in insufficient learning of user interests.

Inspired by the success of Swin Transformer~\cite{swin} in computer vision, we propose a shifted chunk-based self-attention (SC-SA) to overcome the aforementioned limitations.
The schema of SC-SA is illustrated in \cref{fig:sa}(c), and its core design is the shift of chunks to allow for cross-chunk connections while maintaining efficient.
Specifically, every two successive transformer blocks contain a C-SA and an SC-SA (as shown in \cref{fig:arch}).
The first block performs C-SA.
Then, the next block cyclically shifts the chunks by half a chunk size, performs self-attention in each new chunk to aggregate cross-chunk information, and finally reverses the cyclic shift to restore the original chunk partition.
The successive shifted chunk-based transformer blocks can be formulated as:

\begin{equation}
\begin{aligned}
    & a_i = \text{C-SA}(\text{LN}(b_{i-1})) + b_{i-1} \\
    & b_i = \text{MLP}(\text{LN}(a_i)) + a_i \\
    & a_{i+1} = \text{SC-SA}(\text{LN}(b_i)) + b_i \\
    & b_{i+1} = \text{MLP}(\text{LN}(a_{i+1})) + a_{i+1}
\end{aligned}
\end{equation}
where $a_i$ and $b_i$ are the ouputs of self-attention and MLP in the $i$-th block, respectively.

For each chunk containing $c$ embeddings, self-attention is formulated as:
\begin{equation}
\text{Self-Attention}(Q, K, V) = \text{soft} \text{max} (Q K^{\mathsf{T}} / \sqrt{d} + M) V
\end{equation}
where $Q, K, V\in \mathbb{R}^{c \times d}$ are the transformed query, key and value matrices in the chunk, $d$ is feature dimension.
We prohibit communication between embeddings from different buckets, and use a mask $M\in \mathbb{R}^{c\times c}$ to indicate whether the query/key pairs come from the same bucket:
\begin{equation}
 M_{i,j} = \left\{
 \begin{aligned}
 0,\quad & \text{if query } i \text{ and key } j \text{ come from the same bucket}\\
 -\infty,\quad & \text{otherwise}
 \end{aligned}
 \right.
\end{equation}

\subsection{Target Attention}
We have explored the internal correlations among similar embeddings by $T$ shifted chunk-based transformer blocks.
Next, we perform average pooling in each chunk to obtain $C$ interest vectors $\{ x^i_1, x^i_2, \dots, x^i_C\}$ that represent user diverse interests, where $C = L/c$ denote the number of chunks.
Each interest vector is dynamically activated w.r.t. the target item:
\begin{equation}
    s_j = \text{MLP}_2(x^i_j \Vert \text{MLP}_1(x^t))
\end{equation}
where $\text{MLP}_1$ is used for feature space transformation and dimension matching, $\text{MLP}_2$ outputs an attention score that measures the relevance between the target item and the $j$-th interest vector, and $\Vert$ denotes concatenation.

Finally, the user's interest representation towards the target item is calculated as:
\begin{equation}
    x^i = \sum_{j=1}^C s_j x^i_j 
\end{equation}

\subsection{Optimize Objective}
We concatenate the extracted user interest representation $x^i$ with the target item embedding $x^t$ and user profile embedding $x^u$, and input the resulting features into an MLP to predict CTR:
\begin{equation}
\hat{y} = \text{sigmoid} (\text{MLP} (x^i \Vert x^t \Vert x^u))
\end{equation}
The model is optimized with binary cross entropy loss (BCE loss):
\begin{equation}
\label{eq:loss_ctr}
\ell=-\frac{1}{|\mathcal{D}|} \sum_{\mathcal{D}}(y \log \hat{y}+(1-y) \log (1-\hat{y}))
\end{equation}
\label{sec:objective}
where $\mathcal{D}$ is the training set and $y \in \{0,1\}$ is the click label.
\section{Experiments}
\subsection{Experimental Settings}

\subsubsection{Dataset}

We build an offline experimental dataset, termed MeituanCTR, based on the service log of Meituan Waimai\footnote{One of the largest food recommendation platforms in the world.} recommendation advertisement.
We collect and sample one month of data from 2023-04-15 to 2023-05-15 as the training set, data on 2023-05-16 as the validation set, and data on 2023-05-17 as the test set.
This data contains 0.21 billion users and 9.3 billion samples.

\subsubsection{Evaluation Metrics}
We adopt LogLoss in \cref{eq:loss_ctr} and Area Under Curve (AUC) as the evaluation metrics to assess model performance.
 The LogLoss metric measures the gap between the model's predicted results and the labels, with smaller values indicating better performance.
The AUC metric measures the probability of a model to rank a randomly selected positive instance higher than a randomly selected negative one,
with larger values indicating better performance.
The formula for AUC is:
\begin{equation}
\begin{aligned}
\label{eq:AUC_RelaImpr}
 \text{AUC} =
 \frac{1}{|\mathbb{D}^{+}| |\mathbb{D}^{-}|}
 \sum_{x^{+} \in \mathbb{D}^{+}} \sum_{x^{-} \in \mathbb{D}^{-}}
 (\mathbb{I} (f (x^{+}) > f (x^{-}))) \\
\end{aligned}
\end{equation}
where $\mathbb{D}^{+}$ and $\mathbb{D}^{-}$ represent the positive and negative samples, respectively. $\mathbb{I}$ denotes the indicator function, and $f(\cdot)$ is the CTR prediction function.

\subsection{Offline Experiments}
\subsubsection{Compared Methods}
We compare TBIN with the following methods that extract user interest from user behavior data:
\begin{itemize}[leftmargin=*]
    \item DIN~\cite{din} innovatively extracts user interests from their behavior data and has been successfully applied in industry.
    
    \item DIEN~\cite{dien} adopts GRU~\cite{gru} to dynamically updates user interests based on their behavior data over time.
    
    \item SIM~\cite{sim} uses a search-based user interest modeling approach that utilizes lifelong sequential behavior data for CTR prediction.
    
    \item IDA-SR~\cite{idasr} utilizes a pre-trained language model to learn user interest representations from textual features instead of relying on ID-based features (e.g., item ID).
    
    \item UniSRec~\cite{unis_rec} uses item description text to learn universal item and sequence representations to implement transferable representations across different recommendation scenarios.
\end{itemize}
Note that DIN/DIEN/SIM only take ID-based features as input, while IDA-SR/UniSRec/TBIN utilize both ID-based and textual features.
For fairness, we demonstrate the performance of TBIN with only ID-based features in ablation studies.

\subsection{Performance Comparison}
The offline experimental results are summarized in \cref{tab:performance}, where TBIN performs the best.
DIN/DIEN/SIM have poorer performance due to the lack of utilization of textual information, and the carefully designed architecture of TBIN helps it outperform other textual data-based methods (IDA-SR/UniSRec) by a large margin.

\begin{table}[h]
\centering
  \caption{Offline experimental results.
  $\downarrow$ and $\uparrow$ indicate lower-the-better and higher-the-better, respectively.
  The results are presented in the form of mean $\pm$ standard deviation.
  }
  \setlength{\tabcolsep}{1.0mm}{
  \begin{tabular}{l|c|c}
    \toprule
    \text{Methods} & \text{LogLoss}$\downarrow$ &  \text{AUC}$\uparrow$ \\
    \midrule
     DIN & 0.1830$\pm$ 0.0002 & 0.6953 $\pm$ 0.0003 \\
     DIEN & 0.1828$\pm$ 0.0002 & 0.6978$\pm$ 0.0003 \\
     SIM  & 0.1821$\pm$ 0.0003 & 0.7037$\pm$ 0.0004 \\
    \midrule
     IDA-SR & 0.1815$\pm$ 0.0002 & 0.7061$\pm$ 0.0003 \\
     UniSRec & 0.1814$\pm$ 0.0003 & 0.7066$\pm$ 0.0004 \\
    \midrule
    TBIN & 0.1810$\pm$ 0.0002 & 0.7114 $\pm$ 0.0003 \\
    \bottomrule
  \end{tabular}}
\label{tab:performance}
\end{table}

\subsection{Ablation Studies}
\cref{tab:ablation} presents several ablation studies for verifying designs in TBIN.
In TBIN selected as the baseline, the behavior data has a length of 1000 and the number of transformer blocks is set as $T=4$.
\begin{itemize}[leftmargin=*]
    \item \textbf{Importance of textual behavior data.}
    We train a TBIN using only ID-based behavior data (termed as TBIN-ID), and the performance loss indicates that textual behavior data contains rich user interest information.
    
    \item \textbf{Impact of behavior data length.}
    We train TBIN models using behavior data of different lengths. TBIN-100 and TBIN-500 use behavior data with lengths of 100 and 500, respectively. Both models perform worse than the baseline, indicating that longer behavior data helps the model better understand user interest.

    \item \textbf{Impact of the number of transformer blocks.}
    The baseline model uses 4 transformer blocks and performs better than TBIN-T2, which only has 2 blocks, because it is deeper and could learn feature interactions better.
    TBIN-T6, which contains 6 blocks, performs slightly better than the baseline.
    Considering efficiency, TBIN ultimately only uses 4 transformer blocks.

    \item \textbf{Impact of self-attention used in transformer block.}
    We train two TBIN models by replacing all self-attentions in the transformer blocks with chunk-based self-attention (C-SA) and global self-attention (G-SA, where all embeddings attend to each other), respectively.
    Both models perform worse than the baseline.
    The former is due to the lack of information interaction between correlated chunks, while the latter is attributed to the introduction of too much noise by G-SA, in our opinion.
    
\end{itemize}

\begin{table}[h]
\centering
  \caption{Ablation studies of TBIN designs.}
  \setlength{\tabcolsep}{1.0mm}{
  \begin{tabular}{l|c|c}
    \toprule
    \text{Methods} & \text{AUC}$\uparrow$ & $\Delta$\text{AUC} \\
    \midrule
    TBIN (Baseline)& \textbf{0.7114} & \textbf{-}\\
    TBIN-ID & 0.7059 & -0.0055\\
    \midrule
    TBIN-100 & 0.7067 & -0.0047\\ 
    TBIN-500 & 0.7088 & -0.0026\\
    \midrule
    TBIN-T2 & 0.7082 &  -0.0032 \\
    TBIN-T6 & 0.7119 &  +0.0005 \\
    \midrule
    TBIN-C-SA & 0.7094 & -0.0020 \\
    TBIN-G-SA & 0.7089 & -0.0025 \\
    \bottomrule
  \end{tabular}}
\label{tab:ablation}
\end{table}

\subsection{Online A/B Test}
To verify TBIN's performance in real-world scenario, we deploy it on the online advertising system of Meituan Waimai with 10\% traffic volume for one week.
Compared to a long-term optimized baseline, TBIN achieves a significant improvement of +2.7\% in CTR, +2.4\% in Cost Per Mille (CPM), and +1.2\% in Gross Merchandise Volume (GMV), highlighting its importance for generating revenue on the platform.
\section{Conclusion}
This paper proposes a Textual Behavior-based Interest Chunking Network (TBIN) for CTR prediction.
Both offline and online experiments demonstrate the effectiveness of TBIN.

\bibliographystyle{ACM-Reference-Format}
\bibliography{ref}

\end{document}